\documentclass[pdflatex,sn-mathphys-num,iicol]{sn-jnl}

\usepackage{graphicx}%
\usepackage{multirow}%
\usepackage{amsmath,amssymb,amsfonts}%
\usepackage{amsthm}%
\usepackage{mathrsfs}%
\usepackage[title]{appendix}%
\usepackage{xcolor}%
\usepackage{textcomp}%
\usepackage{manyfoot}%
\usepackage{booktabs}%
\usepackage{algorithm}%
\usepackage{algorithmicx}%
\usepackage{algpseudocode}%
\usepackage{listings}%
\usepackage{ulem}

\theoremstyle{thmstyleone}%
\theoremstyle{thmstyletwo}%

\theoremstyle{thmstylethree}%

\usepackage{bm}
\usepackage{physics}

\newcommand{\kpc}  {\,\text{kpc}}  
\newcommand{\fm}   {\,\text{fm}}   
\newcommand{\keV}  {\,\text{keV}}  
\newcommand{\MeV}  {\,\text{MeV}}  
\newcommand{\GeV}  {\,\text{GeV}}  

\begin{document}

\title[Measuring the Low-Energy Weak Mixing Angle with Supernova Neutrinos]{Measuring the Low-Energy Weak Mixing Angle with Supernova Neutrinos}


\author[1]{Chun-Ming Yip
\href{https://orcid.org/0000-0002-3311-5387}{\includegraphics[scale=0.05]{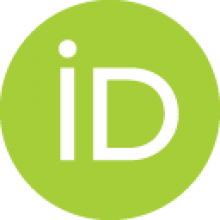}}}\email{cmyip1999@gmail.com}

\author*[1]{Xu-Run Huang
\href{https://orcid.org/0000-0003-1842-8657}{\includegraphics[scale=0.05]{orcid-ID.png}}}\email{bryce.xr.huang@gmail.com}

\author[1]{Ming-chung Chu
\href{https://orcid.org/0000-0002-1971-0403}{\includegraphics[scale=0.05]{orcid-ID.png}}}\email{mcchu@phy.cuhk.edu.hk}

\author[1]{Qishan Liu
\href{https://orcid.org/0000-0003-1437-6829}{\includegraphics[scale=0.05]{orcid-ID.png}}}\email{qisliu@link.cuhk.edu.hk}

\affil[1]{Department of Physics and Institute of Theoretical Physics, The Chinese University of Hong Kong, Shatin, N.T., Hong Kong S.A.R., People's Republic of China}


\abstract{
The weak mixing angle $\theta_W$ is a fundamental parameter in the electroweak theory with a value running according to the energy scale, and its precision measurement in the low-energy regime is still ongoing. 
We propose a method to measure the low-energy $\sin{^2\theta_W}$ by taking advantage of {Argo}, a future ton-scale liquid argon dark matter detector, and the neutrino flux from a nearby core-collapse supernova (CCSN). 
We evaluate the expected precision of this measurement through the coherent elastic neutrino-nucleus scattering (CE$\nu$NS) channel. 
We show that {Argo} is potentially capable of achieving a few percent determination of $\sin{^2\theta_W}$, at the momentum transfer of $q\!\sim\!20{\MeV}$, in the observation of a CCSN within $\sim\!3{\kpc}$ from the Earth. 
Such a measurement is valuable for both the precision test of the electroweak theory and searching for new physics beyond the standard model in the neutrino sector.
}



\maketitle

\section{Introduction} \label{Section 1}

The weak mixing angle $\theta_W$ is one of the most fundamental parameters in the Standard Model (SM) of particle physics, which characterizes the mixing of vector bosons from the $SU\left(2\right)_L \times U\left(1\right)_Y$ symmetry~\cite{Weinberg:1967tq}. 
Collider experiments in the last decades had achieved high precision measurements of $\sin{^2\theta_W}$ near the so-called 'electroweak scale' of $\sim\!100{\GeV}$~\cite{ParticleDataGroup:2022pth}. 
The value of $\sin{^2\theta_W}$ runs along with the energy scale where it is measured due to quantum effects, particularly at the next-to-leading-order.
The SM prediction at zero momentum transfer $q=0$ is $\sin{^2\theta_W}=0.23863\pm0.00005$ under the modified minimal subtraction ($\overline{MS}$) renormalization scheme~\cite{ParticleDataGroup:2022pth}. 
Experiments are also performed to check this prediction.
Nevertheless, the precise determination of $\sin{^2\theta_W}$ in the low-energy regime is still an ongoing issue~\cite{Kumar:2013yoa}.

Previous efforts can be categorized into the charged lepton and neutrino sectors.
In the former, the {Qweak} Collaboration recently reported its result with a precision of $\pm0.5\%$ at $q=0.158{\GeV}$ via the measurement of parity-violation asymmetry in a polarized electron scattering experiment~\cite{Qweak:2018tjf}, which agrees well with the SM prediction.
In the low-energy regime, the most precise results to date are provided at $q\simeq2.4{\MeV}$ by the atomic parity violation (APV) experiments, but they turn out to be smaller than the SM prediction by $\sim\!1.5\sigma$~\cite{Dzuba:2012kx,Roberts:2014bka}.
In the neutrino sector, an experimental anomaly ($3\sigma$) is reported by the {NuTeV} experiment, concerning the SM prediction for neutrino-nucleon scatterings at $q\!\sim\!1{\GeV}$~\cite{NuTeV:2001whx}.
At the MeV scale, the coherent elastic neutrino-nucleus scattering (CE$\nu$NS) is a powerful probe, which was predicted theoretically in 1973~\cite{Freedman:1973yd,Freedman:1977xn} but only observed experimentally for the first time in 2017~\cite{COHERENT:2017ipa}.
In this interaction, the recoil of the nucleus is the only observable signature, which is very difficult to detect since it shows up with an extremely small kinetic energy.
Observations have been achieved with neutrinos from stopped pion decays by the COHERENT collaboration using a cesium-iodide (CsI) detector~\cite{COHERENT:2017ipa} and a liquid argon detector~\cite{COHERENT:2020iec}.
The COHERENT-Ge~\cite{COHERENT:2024axu}, Dresden-II~\cite{Colaresi:2022obx} and CONUS+~\cite{Ackermann:2025obx} experiments observed reactor antineutrinos by CE$\nu$NS on germanium. 
The recent detection of solar $^{8}\text{B}$ neutrinos on dark matter (DM) detectors, reported by the PandaX-4T~\cite{PandaX:2024muv} and XENONnT~\cite{XENON:2024ijk} collaborations, are also conducted via the CE$\nu$NS.
Limited by the sparse data, an accurate extraction of $\sin{^2\theta_W}$ still requires further efforts (for recent progress, see Refs.~\cite{Pandey:2023arh,Cadeddu:2023tkp}).
Meanwhile, the nature of neutrinos remains elusive. 
New physics beyond the SM may exist in the low-energy neutrino-nucleon interactions and shift the extracted value of $\sin{^2\theta_W}$ in these interactions~\cite{Barranco:2005yy,Kumar:2013yoa,Billard:2018jnl,Cadeddu:2018dux}, e.g., neutrino non-standard interactions (NSIs)~\cite{Wolfenstein:1977ue,Ohlsson:2012kf,10.21468/SciPostPhysProc.2.001}, neutrino charge properties~\cite{Giunti:2014ixa}, etc.. 
Any experimental constraints in this sector also render valuable information on searching for new physics beyond the SM.

Such a scientific potential stimulates further experimental and theoretical attention to the CE$\nu$NS.
Using low-energy neutrino fluxes from nuclear reactors or stopped-pion decays in accelerators, some terrestrial CE$\nu$NS experiments are running, under construction, or being planned, e.g.,{CONNIE}~\cite{CONNIE:2016ggr}, {MINER}~\cite{MINER:2016igy}, NEON~\cite{NEON:2022hbk}, NUCLEUS~\cite{NUCLEUS:2019igx}, Ricochet~\cite{Colas:2021pxr}, {RED-100}~\cite{Akimov:2022xvr}, and so on~\cite{Baxter:2019mcx}.
Meanwhile, based on some of the above reactor neutrino experiments, the prospect of measuring the low-energy $\sin{^2\theta_W}$ has been discussed, e.g., a sub-{$1\%$} extraction precision is found to be possible in ideal cases~\cite{Canas:2018rng}. 
On the other hand, several determinations of the $\sin{^2\theta_W}$ at the MeV energy scale have already been obtained by analyzing the recent CE$\nu$NS data (e.g., COHERENT, Dresden-II, CONUS+, PandaX-4T and XENONnT, etc.)~\cite{AristizabalSierra:2022axl,AtzoriCorona:2022qrf,Majumdar:2022nby,DeRomeri:2022twg,AtzoriCorona:2023ktl,Maity:2024aji,DeRomeri:2024iaw,Alpizar-Venegas:2025wor,Chattaraj:2025fvx,DeRomeri:2025csu}. The error bars are still large (see Fig.\ref{figure5}). Some efforts have also been made in the context of combining other low-energy probes and the measurement of the neutron radius of $^{133}{\rm Cs}$~\cite{Huang:2024jbh,AtzoriCorona:2024vhj}.


CE$\nu$NS measurements can also be conducted using the low-energy neutrino flux from a galactic core-collapse supernova (CCSN). 
The first-ever supernova neutrino detection was done for SN 1987A, 
by the Kamiokande II, Irvine-Michigan-Brookhaven (IMB), and Baksan detectors~\cite{Kamiokande-II:1987idp,IMB:1987klg,1988SvAL...14...41A}.
Sophisticated CCSN simulations, in the meantime, have been performed by different groups to predict the neutrino signals of CCSNe~\cite{Hudepohl:2013zsj,Mirizzi:2015eza,Nagakura:2020qhb,Burrows:2020qrp}.
It is found that $\sim\!10^{58}$ neutrinos, with mean energies of $\sim\!10-20{\MeV}$, are released within a few seconds after a CCSN explosion occurs. 
Such a neutrino flux can be detected by large-scale DM detectors. 
For example, the sensitivity of future xenon and argon DM detectors to supernova neutrinos has been investigated, and $\gtrsim\!10^2$ CE$\nu$NS events are expected to be observed for a typical CCSN at $10{\kpc}$ from the Earth~\cite{Lang:2016zhv,DarkSide20k:2020ymr}.
Meanwhile, other dedicated large-scale neutrino observatories, such as Super-Kamiokande~\cite{Super-Kamiokande:2016kji} (Hyper-Kamiokande~\cite{Hyper-Kamiokande:2021frf}), DUNE~\cite{DUNE:2020zfm}, RES-NOVA~\cite{Pattavina:2020cqc,RES-NOVA:2021gqp}, JUNO~\cite{JUNO:2021vlw}, and so on~\cite{Scholberg:2012id,Mirizzi:2015eza,Horiuchi:2018ofe}, are capable of collecting a substantial number of supernova neutrino events, e.g., $\mathcal{O} (10^5)$ for a source at $10{\kpc}$~\cite{Mirizzi:2015eza,Horiuchi:2018ofe}.
So, a combined analysis of data from multiple detectors may potentially achieve a precise reconstruction of supernova neutrino energy spectrum for a nearby CCSN~\cite{Nikrant:2017nya,GalloRosso:2017mdz,Li:2017dbg,Li:2019qxi,Nagakura:2020bbw,Huang:2023aob}. 


Measuring the low-energy $\sin{^2\theta_W}$ is feasible by taking advantage of the reconstructed supernova neutrino spectrum from large-scale neutrino observatories and the CE$\nu$NS events in large-scale DM detectors.
In this work, we estimate the potential of measuring the low-energy $\sin{^2\theta_W}$ with {Argo}, a proposed next-generation liquid argon DM detector~\cite{DarkSide-20k:2017zyg,DarkSide20k:2020ymr}, in the observation of a nearby CCSN. 
The Argo experiment is suitable for this task. It has a large expected active mass of $362.7$ tons~\cite{DarkSide20k:2020ymr}, considerably larger than other large-scale DM detectors (e.g., $43$ tons for {PandaX-xT}~\cite{PANDA-X:2024dlo}, $5.9$ tons for {XENONnT}~\cite{XENON:2024wpa} and $40$ tons for {DARWIN}~\cite{DARWIN:2016hyl}), which renders a high sensitivity to supernova neutrinos. Compared with heavier nuclei (e.g., Xe in other DM detectors), selecting the Ar nucleus can alleviate the uncertainty from the undetermined neutron distribution
in the nucleus. Currently, the \textit{ab initio} calculation of the neutron distribution is easier and more precise for light and medium-mass nuclei (see, e.g., Ref.~\cite{Hagen:2015yea}).
This paper is organized as follows. 
In Sect.~\ref{Section 2}, we introduce the general properties of supernova neutrinos and 4 example fluxes obtained by numerical simulations. 
In Sect.~\ref{Section 3}, we present the detection prospect in {Argo}. 
The expected precision and relevant uncertainty analysis of the low-energy $\sin{^2\theta_W}$ measurement are shown in Sect.~\ref{Section 4}. 
Finally, we conclude in Sect.~\ref{Section 5}. 

\section{Supernova neutrino flux} \label{Section 2}

In a real measurement, the neutrino flux can be determined by observations, including the direct measurements by dedicated neutrino observatories and outcomes from CCSN models which have been calibrated by multi-messenger observations.
In this paper, we use neutrino fluxes from CCSN simulations in the literature.
Thanks to tremendous progress in modeling CCSN phenomena in the last decades, modern hydrodynamic codes are capable of showing robust explosions in simulations~\cite{Muller:2016izw,Burrows:2020qrp}.
The supernova neutrino energy spectrum of one species ($\nu_\beta$ in $\nu_e,\nu_\mu,\nu_\tau$, and anti-particles) is well approximated by a quasi-thermal distribution (the so-called Garching formula~\cite{Keil:2002in,Tamborra:2012ac}) as a function of neutrino energy $E_{\nu_\beta}$:
\begin{equation}
\label{eq:GarchingFormula}
    f_{\nu_\beta} \left( E_{\nu_\beta} \right) = C_\beta \left({\frac{E_{\nu_\beta}}{\langle E_{\nu_\beta} \rangle}}\right)^{\alpha_\beta} \exp{-\left( \alpha_\beta+1 \right) \frac{E_{\nu_\beta}}{\langle E_{\nu_\beta} \rangle}},
\end{equation}
where $\langle E_{\nu_\beta} \rangle$ is the mean neutrino energy, $\alpha_\beta$ is the shape parameter which controls the suppression of high energy tail, and the normalization constant $C_\beta$ is given by
\begin{equation}
    C_\beta = \frac{\left( \alpha_\beta+1 \right)^{\alpha_\beta+1}}{\langle E_{\nu_\beta} \rangle \Gamma\left( \alpha_\beta+1 \right)}
\end{equation}
for the species $\nu_\beta$, with $\Gamma$ denoting the gamma function. 
Neutrino flavor transitions~\cite{Scholberg:2017czd} have no impact on our results since the CE$\nu$NS is flavor-blind. 
The total neutrino flux on the Earth can be evaluated by summing over all neutrino species, i.e.,
\begin{equation}
    \Phi \left( E_\nu \right) = \sum_{\beta} \frac{1}{4 \pi d^2} \frac{E^{tot}_{\nu_\beta}}{\langle E_{\nu_\beta} \rangle} f_{\nu_\beta} \left( E_{\nu_\beta} \right)
\label{eq: SN neutrino flux}
\end{equation}
as a function of neutrino energy $E_\nu$ regardless of species, with $d$ being the distance of the CCSN and $E^{tot}_{\nu_\beta}$ the total energy emitted through one species.
In practice, Eq.~\eqref{eq:GarchingFormula} is used to describe the energy spectrum of both the time-integrated flux as $f_{\nu_\beta} \left( E_{\nu_\beta} \right)$, and the time-dependent flux as $f_{\nu_\beta} \left( E_{\nu_\beta},t \right)$ with $\alpha_{\beta}(t)$ and $\langle E_{\nu_\beta}(t) \rangle$.
Correspondingly, Eq.~\eqref{eq: SN neutrino flux} can also describe the total differential flux $\Phi \left( E_\nu,t \right)$ with $\langle E_{\nu_\beta}(t) \rangle$ and $f_{\nu_\beta} \left( E_{\nu_\beta},t \right)$, after replacing $E^{tot}_{\nu_\beta}$ with the neutrino luminosity $\mathcal{L}_{\nu_\beta} (t)$.
The time-integrated flux is defined as $\Phi \left( E_\nu \right) \equiv \int \Phi \left( E_\nu,t \right) \mathrm{d}t$.

\begin{figure}[!ht]
    \includegraphics[width=\columnwidth]{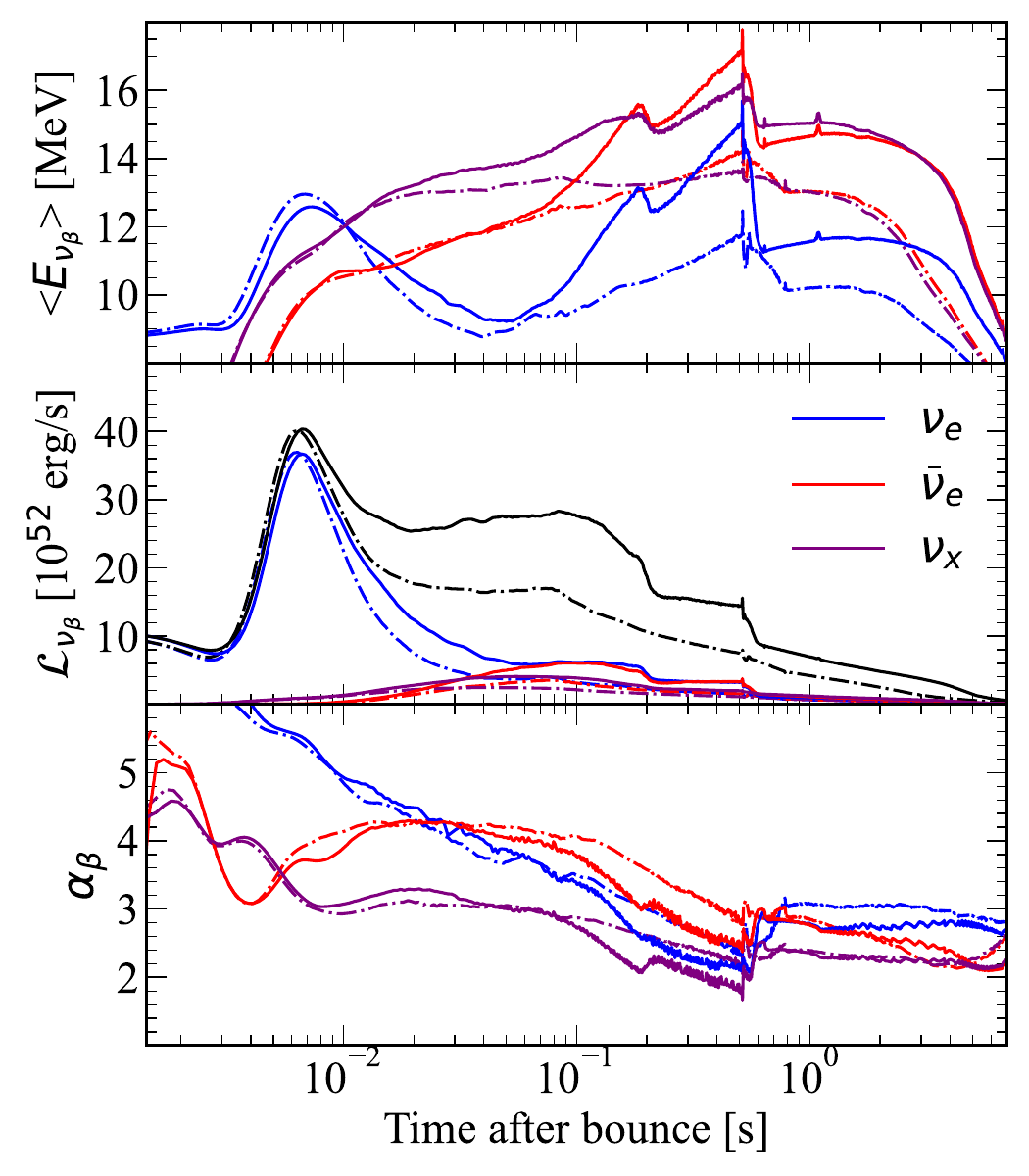}
    \caption{
    Temporal evolution of supernova neutrino spectral parameters for the models s11.2-S (dashed-dotted lines) and s27.0-L (solid lines). 
    $\nu_e$, $\Bar{\nu}_e$, and $\nu_x$ ($\nu_x$ denotes one of $\nu_\mu$, $\Bar{\nu}_\mu$, $\nu_\tau$, and $\Bar{\nu}_\tau$) are colored blue, red, and purple, respectively.
    Upper panel: neutrino mean energies $\langle E_{\nu_\beta} \rangle$. 
    Middle panel: neutrino luminosities $\mathcal{L}_{\nu_\beta}$. 
    The total neutrino luminosities are also shown in black. 
    Lower panel: shape parameter $\alpha_\beta$.
    Note that the feature appearing in all curves at $\sim\!0.5\,\rm s$ is due to the artificial triggering of the explosion models (see more discussion in the text)~\cite{Mirizzi:2015eza}
    }
    \label{figure1}
\end{figure}

\begin{table}[!ht]
    \caption{
    \label{tab: CCSN models}
    The CCSN models from the Garching group~\cite{Hudepohl:2013zsj,Mirizzi:2015eza}. 
    The progenitor mass and nuclear EoS of each model are listed
    }
    \begin{tabular}{ccc}
    \toprule
    Model & Progenitor mass [M$_\odot$] & EoS \\
    \midrule
    s11.2-S & 11.2 & Shen  \\
    s11.2-L & 11.2 & LS220 \\
    s27.0-S & 27.0 & Shen  \\
    s27.0-L & 27.0 & LS220 \\
    \botrule
    \end{tabular}
\end{table}
The neutrino fluxes from CCSN simulations depend not only on the properties of progenitor star (e.g., mass, compactness, etc.~\cite{OConnor:2012bsj,Seadrow:2018ftp}) but also on the nuclear matter equation of state (EoS)~\cite{Pan:2017tpk,daSilvaSchneider:2020ddu,Raduta:2021coc}. 
We use realistic values of $\alpha_\beta (t)$, $\langle E_{\nu_\beta} (t) \rangle$ and $\mathcal{L}_{\nu_\beta} (t)$ from 4 CCSN explosion models provided by the Garching group, considering the above factors.
These models are listed in Table~\ref{tab: CCSN models},
and refer to Refs.~\cite{Hudepohl:2013zsj,Mirizzi:2015eza}.
The temporal evolution of the neutrino spectral parameters is shown in Fig.~\ref{figure1} for the models s11.2-S and s27.0-L, where the former (latter) model results in the smallest (largest) expected number of event counts among the 4 models (see Sect.~\ref{Section 3}).
The explosions of these CCSN models were artificially triggered at $\sim\!0.5\,\rm s$ after bounce and some kinks appear accordingly in the neutrino curves at $\sim\!0.5\,\rm s$, which do not lead to any significant perturbing consequences for the subsequent proto-neutron star cooling evolution~\cite{Mirizzi:2015eza}.
The feature seen during this time will not appear in self-consistent explosion models. 
We find that the neutrinos emitted during $0.5\!\sim\!0.53\,\rm s$ only contribute to the total CE$\nu$NS events by $<\!2\%$.
Note that the time-integrated flux $\Phi \left( E_\nu \right)$ is already sufficient to compute the total number of CE$\nu$NS events, and that is also what we can directly construct from the detection by other neutrino observatories.

\section{Detection prospect in Argo} \label{Section 3}

The primary purpose of conventional direct DM detectors is to search for nuclear recoils induced by galactic DM particles (e.g., the weakly interacting massive particles). 
The CE$\nu$NS signals induced by solar neutrinos, reactor neutrinos, atmospheric neutrinos, the diffuse supernova neutrino background, and geoneutrinos will become an unavoidable background as the sensitivity of DM detectors approaches keV level~\cite{MarrodanUndagoitia:2015veg,OHare:2021utq}.
Conversely, these detectors can act as excellent low-energy neutrino observatories. 
The opportunity of observing supernova neutrinos has been investigated for large-scale DM detectors, e.g., liquid xenon detectors like XENONnT, DARWEN, and LZ~\cite{Lang:2016zhv,Khaitan:2018wnf} as well as liquid argon detectors like DarkSide-20k and {Argo}~\cite{DarkSide20k:2020ymr}. In this section, we demonstrate how to obtain the expected CE$\nu$NS event counts of the {Argo} experiment.

The {Argo} experiment is proposed as an extension of DarkSide-20k by the DarkSide Collaboration~\cite{DarkSide-20k:2017zyg,DarkSide20k:2020ymr}. 
It is assumed to have a dual-phase Liquid Argon Time Projection Chamber (LAr TPC) with an active mass of $362.7$ tons~\cite{DarkSide20k:2020ymr}. 
Such a dual-phase TPC is primarily filled with liquid argon and has a gaseous argon phase on the top.
When a CE$\nu$NS event occurs in the instrumented volume, two measurable signals are produced, i.e., a prompt light pulse (S1) induced by scintillation in the liquid phase, and a delayed pulse (S2) associated with ionization electrons. 
The S2-only mode can achieve a lower energy threshold and a better detection efficiency than the S1-only or S1-S2 coincident modes for supernova neutrino detection~\cite{Lang:2016zhv,DarkSide20k:2020ymr}.

The differential cross section of CE$\nu$NS in the SM, as a function of the neutrino energy $E_\nu$ and the nuclear recoil energy $T$, is given by~\cite{Drukier:1984vhf,Lindner:2016wff,Huang:2022wqu,Tomalak:2020zfh}
\begin{equation}
    \dv{\sigma}{T} \left( E_\nu, T \right) =  \frac{G_F^2 M}{4 \pi} Q_W^2 F_W^2 \left( q \right) \left[ 1 - \frac{T}{E_\nu} - \frac{M T}{2 E_\nu^2} \right],
\label{eq: CEvNS}
\end{equation}
where $G_F$ is the Fermi coupling constant, $M$ the mass of the target nucleus with $Z$ ($N$) protons (neutrons), $Q_W$ the weak charge of the target nucleus, and $F_W \left( q \right)$ the corresponding weak form factor. 
The momentum transfer $q$ is given by $q^2 = 2 M E_\nu^2 T / (E_\nu^2 - E_\nu T) \simeq 2 M T$ for $E_\nu \gg T$.
Note that Eq.~\ref{eq: CEvNS} describes the interaction between a neutrino and a nucleus with spin-0~\cite{Lindner:2016wff}.
Natural argon is $99.6\%$ $^{40}\mathrm{Ar}$ with spin-0~\cite{Meija:2016pac}, so we neglect the contributions from other isotopes.
The mass of $^{40}\mathrm{Ar}$ can be calculated via $M = Z \times m_p + N \times m_n - \left( Z + N \right) E_B$, with $m_p$ ($m_n$) being the rest mass of proton (neutron) and $E_B = 8.595259{\MeV}$ the binding energy per nucleon~\cite{Wang:2021xhn}. 

The weak charge of a nucleus $Q_W$ is given by
\begin{equation}
    Q_W = -2 \left( Z g_V^p + N g_V^n \right),
\end{equation}
where the proton and neutron vector couplings are $g_V^p = 1/2 - 2 \sin{^2\theta_W}$ and $ g_V^n = -1/2$, respectively, at the tree level. 
The weak form factor $F_W \left( q \right)$ can be computed by
\begin{equation}
    F_W \left( q \right) = -2 \frac{\left[ Z g_V^p F_p \left( q \right) + N g_V^n F_n \left( q \right) \right]}{Q_W},
\end{equation}
where $F_{n(p)} \left( q \right)$ is the neutron (proton) form factor. 
We adopt the Helm parametrization~\cite{Helm:1956zz,Piekarewicz:2016vbn} since it is very successful in analyzing electron scattering form factors~\cite{PhysRev.163.927,Raphael:1970yd,Friedrich:1982esq} and has also been used in previous CE$\nu$NS studies~\cite{Cadeddu:2017etk,Huang:2019ene,Cadeddu:2020lky}. 
It has the form as
\begin{equation}
    F_i \left( q \right) = 3 \frac{j_1 \left( q R_{0,i} \right)}{q R_{0,i}} e^{-q^2 s^2 / 2},
\label{eq: Helm parametrization}
\end{equation}
where $i=p,n$ labels the form factor associated with proton and neutron, respectively, and $j_1 \left( x \right) = \sin{x}/x^2 - \cos{x}/x$ is the spherical Bessel function of order one.
The diffraction radius $R_{0,i}$ is related to the root-mean-squared (rms) radius $R_i$ by
\begin{equation}
    R_i^2 = \frac{3}{5} R_{0,i}^2 + 3 s^2,
\end{equation}
with $s$ quantifying the surface thickness. We adopt the value $s=0.9 \fm$, following Ref.~\cite{Cadeddu:2020lky}. 
$R_{i}$ is associated with the point-nucleon distribution radius $R_{i}^{\text{point}}$ and the radius of nucleon $\langle r_{i}^2 \rangle^{1/2}$, i.e.,
\begin{equation}
    R_i^2 = (R_{i}^{\text{point}})^2 + \langle r_{i}^2 \rangle.
\end{equation}
The radius of a proton equals its charge radius and has the value of $\langle r_p^2 \rangle^{1/2} = 0.8414(19) \fm$~\cite{Hammer:2019uab}.
For neutron, we assume $\langle r_n^2 \rangle^{1/2} \simeq \langle r_p^2 \rangle^{1/2}$, which is supported by the neutron magnetic radius measurement~\cite{ParticleDataGroup:2018ovx}.
Electron scattering experiments have made precise measurements on the charge radius of a nucleus ($R_c$), which is given by~\cite{Ong:2010gf,Horowitz:2012tj}
\begin{equation}
    R_c^2 = (R_{p}^{\text{point}})^2 + \langle r_{p}^2 \rangle + \frac{N}{Z} \langle r_n^2 \rangle_c,
\end{equation}
with $\langle r_n^2 \rangle_c = -0.1161(22) \fm^2$~\cite{ParticleDataGroup:2018ovx} being the squared charge radius of a neutron.
For $^{40}\text{Ar}$, it is measured as $R_c = 3.4274\pm0.0026${\fm}~\cite{Fricke:1995zz,Angeli:2013epw}. 
As a consequence, we obtain the value of $R_p$ as
\begin{equation}
    R_p = 3.448 \pm 0.003 \fm.
\end{equation}
On the other hand, the neutron rms radius is experimentally unknown~\cite{Thiel:2019tkm}, and different nuclear models would give different predictions (see, e.g., Table I in Ref.~\cite{Cadeddu:2020lky}). 
Theoretical predictions indicate that $R_n^{\text{point}}$ is universally larger than $R_p^{\text{point}}$ by $0.08-0.11 \fm$ for $^{40}\text{Ar}$~\cite{Cadeddu:2020lky}. 
We take such an assumption:
\begin{equation}
    R_n \simeq R_p + 0.1 \fm.
\end{equation}
It should be further pointed out that the neutron form factors at $\mathcal{O} (10){\MeV}$ are not sensitive to uncertainties of neutron distributions in nuclei since its value is close to $1$, especially for light nuclei like $^{40}\text{Ar}$ (see, e.g., Fig. 1 in Ref.\cite{Kerman:2016jqp}).
We check it by varying $R_n-R_p$ by $\pm 20\%$ and the resulting CE$\nu$NS event counts only deviate by about $\mp 0.1\%$ for the 4 aforementioned CCSN models.
Also, the choice of the form factor parametrization
could hardly affect the results here, e.g., the symmetrized Fermi and Helm form factors give the same results in the analysis of the COHERENT data~\cite{Cadeddu:2017etk}.

We estimate the number of expected events $\mathcal{N}$ in {Argo} by the expression,
\begin{equation}
    \mathcal{N} = N_{\text{Ar}} \int \mathrm{d}E_\nu \mathcal{E} \left( E_\nu\right) \int_{T_{\text{min}}}^{T_{\text{max}}} \mathrm{d}T \dv{\sigma}{T} \left( E_\nu, T \right) \Phi\left( E_\nu \right).
    \label{eq: expected event counts of CEvNS}
\end{equation}
Here, $N_{\text{Ar}} = N_A m_{\text{det}} / M_{\text{Ar}}$ denotes the number of Ar atoms contained in the active volume of the detector, with $N_A$ being the Avogadro constant, $m_{\text{det}}$ the active mass of the detector, and $M_{\text{Ar}} = 39.96\,\text{g/mol}$ the molar mass of argon.
We consider the $E_\nu$-dependent detection efficiency $\mathcal{E}\left( E_\nu \right)$, which is evaluated in the range of [3, 100] ionization electrons (see Fig. 4 in Ref.~\cite{DarkSide20k:2020ymr}).
This detection efficiency is preferred in the study of supernova neutrino detection by DarkSide-20k and Argo, to suppress the single-electron background and the $^{39}\text{Ar}$ events~\cite{DarkSide20k:2020ymr}.
When integrating over $T$, an energy threshold of $T_{\text{min}} = 0.46${\keV} is considered, since it has been proven as feasible by {DarkSide-50}~\cite{DarkSide:2018bpj} and also applied for {Argo} in Ref.~\cite{DarkSide20k:2020ymr}. 
For the scattering between a Ar nucleus and an incoming neutrino with an energy $E_\nu$, $T_{\text{max}} = 2 E_{\nu}^2 / (M_{\text{Ar}} + 2 E_\nu)$.
The integration over $E_\nu$ starts from $1{\MeV}$, where $\mathcal{E}(E_\nu)$ already decreases to zero, and ends at $60{\MeV}$, which is high enough for the energy spectrum (Eq.~\ref{eq:GarchingFormula}) to almost fade out.
The scattering rate $\mathrm{d}\mathcal{N}/\mathrm{d}t$ can also be computed by Eq.~\eqref{eq: expected event counts of CEvNS} if we replace $\Phi \left( E_\nu \right)$ with $\Phi \left( E_\nu,t \right)$.

\begin{figure}[!ht]
    \includegraphics[width=\columnwidth]{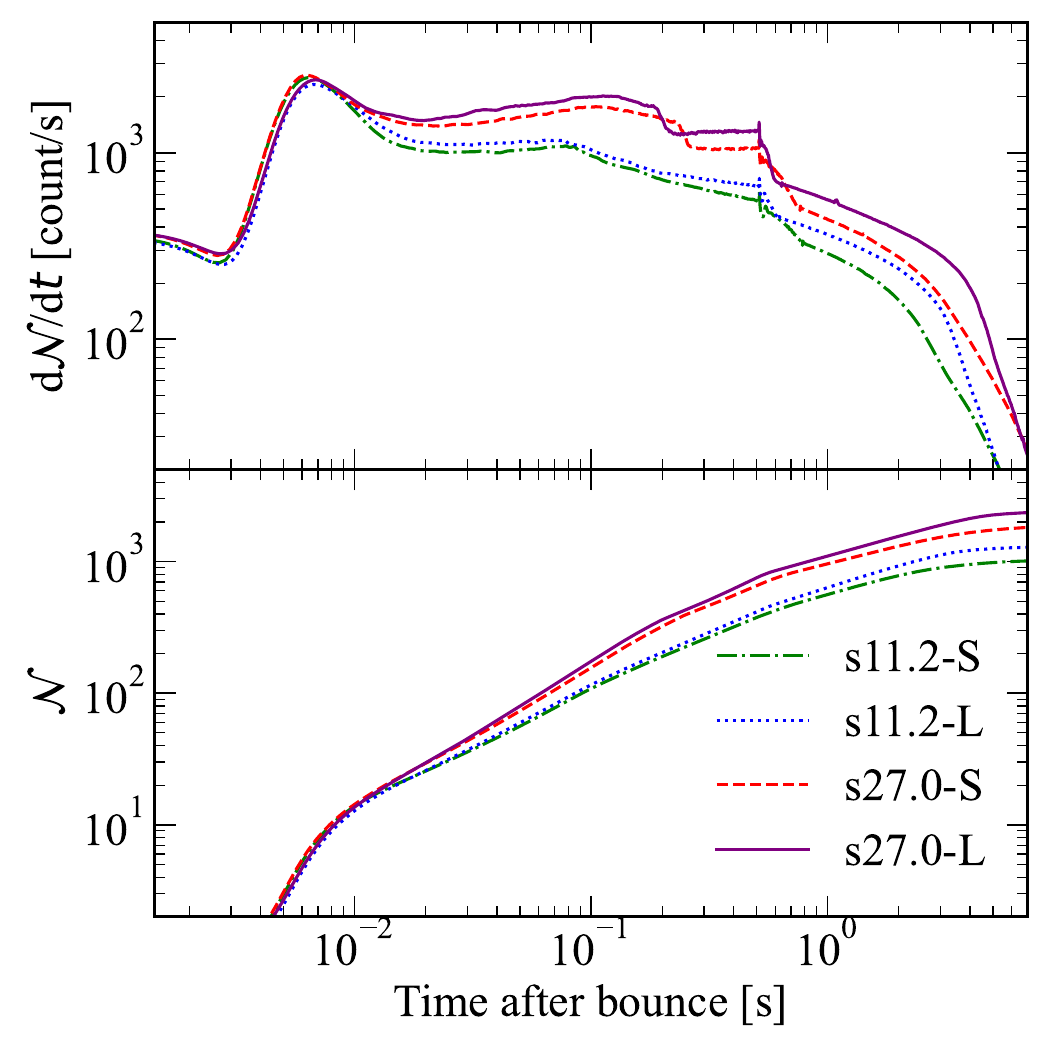}
    \caption{
    Expected event counts of CE$\nu$NS in {Argo} for the 4 models, evaluated at $d=10{\kpc}$. 
    Upper panel: scattering rate $\mathrm{d}\mathcal{N}/\mathrm{d}t$.
    Lower panel: cumulative event counts $\mathcal{N}$ over time
    }
    \label{figure2}
\end{figure}
Figure~\ref{figure2} shows the scattering rate ($\mathrm{d}\mathcal{N}/\mathrm{d}t$) and the corresponding cumulative event counts for the 4 CCSN models, with the $\sin{^2\theta_W}$ in the SM at $d=10{\kpc}$. 
The cumulative event counts $\mathcal{N}$ are calculated to be 1012, 1284, 1820, and 2357 for the 4 CCSN models presented in Table~\ref{tab: CCSN models}, respectively, at $7\,\text{s}$ after bounce and saturate afterward. 
More events are recorded for the models s27.0-S and s27.0-L with heavier progenitors.
We also find that about half of the events are recorded within the first second after the bounce.

\section{Low-energy weak mixing angle sensitivity} \label{Section 4}

\subsection{Statistical analysis} \label{Section 4.1}

To assess the precision of measuring $\sin^2{\theta_W}$, we perform a $\chi^2$ analysis, assuming that the number of observed events by {Argo} is $\mathcal{N}_\text{obs} = \mathcal{N}_\text{true} - \mathcal{N}_\text{bias}$, where $\mathcal{N}_\text{true}$ is the actual number of total CE$\nu$NS interactions and $\mathcal{N}_\text{bias}$ is the number of unresolved interactions due to event pile-up (see later contents for more detailed discussions). 
Following Refs.~\cite{Canas:2018rng,Huang:2022wqu}, we consider the chi-squared function
\begin{equation}
    \chi^2 = \frac{\left( \mathcal{N}_{\text{obs}}^{\text{SM}} - \mathcal{N}_{\text{obs}}^{\text{th}} \right)^2}{\sigma^2_{\text{stat}} + \sigma^2_{\text{syst}}}.
    \label{eq: chi-2 function}
\end{equation}
Here, $\mathcal{N}_{\text{obs}}^{\text{SM}}$ denotes the number of observed events computed with the SM value of $\sin^2{\theta_W}$ at zero momentum transfer, i.e., $\sin^2{\theta_W}=0.23863$~\cite{ParticleDataGroup:2022pth}, while $\mathcal{N}_{\text{obs}}^{\text{th}}$ varies according to the value of $\sin{^2\theta_W}$.
We divide the uncertainty estimation into two parts, i.e., a pure statistical uncertainty $\sigma_\text{stat} = \sqrt{\mathcal{N}_{\text{obs}}^{\text{SM}}}$ and a systematic error $\sigma_{\text{syst}}$. 
In this section, we attempt to draw an estimation based on the current limited knowledge of $\sigma_\text{syst}$.
Firstly, due to the ultra-high event rate and limitation from the time resolution of LAr TPC, the event pile-up will become inevitable for a nearby target. This will reduce the total count collected in {Argo} and introduce a systematic uncertainty.
Secondly, the energy spectra of supernova neutrinos are measured with an uncertainty, and it introduces an uncertainty on the expected number of events in {Argo}. This uncertainty will be more pronounced for a far target. 
The total systematic error of the low-energy $\sin^2{\theta_W}$ measurement considering the above two types of uncertainties can be written as $\sigma^2_{\text{syst}} = \sigma^2_{\text{bias}} + \sigma^2_{\text{flux}}$. 
Here, $\sigma_{\text{bias}}$ denotes the uncertainty in event pile-up estimation in {Argo}; $\sigma_{\text{flux}}$ denotes the flux uncertainty of expected events in {Argo}, which originates from the uncertainty of the extracted supernova neutrino flux $\sigma_{\nu}$ in other detectors. 
Nonetheless, there may also exist more uncertainties from other sources, e.g., the high-order corrections to the cross-section, the configuration of the {Argo} detector, and so on.
In this section, we only consider the previous two types of uncertainties and leave a comprehensive uncertainty evaluation to a real measurement in the future.

\begin{figure}[!ht]
    \includegraphics[width=\columnwidth]{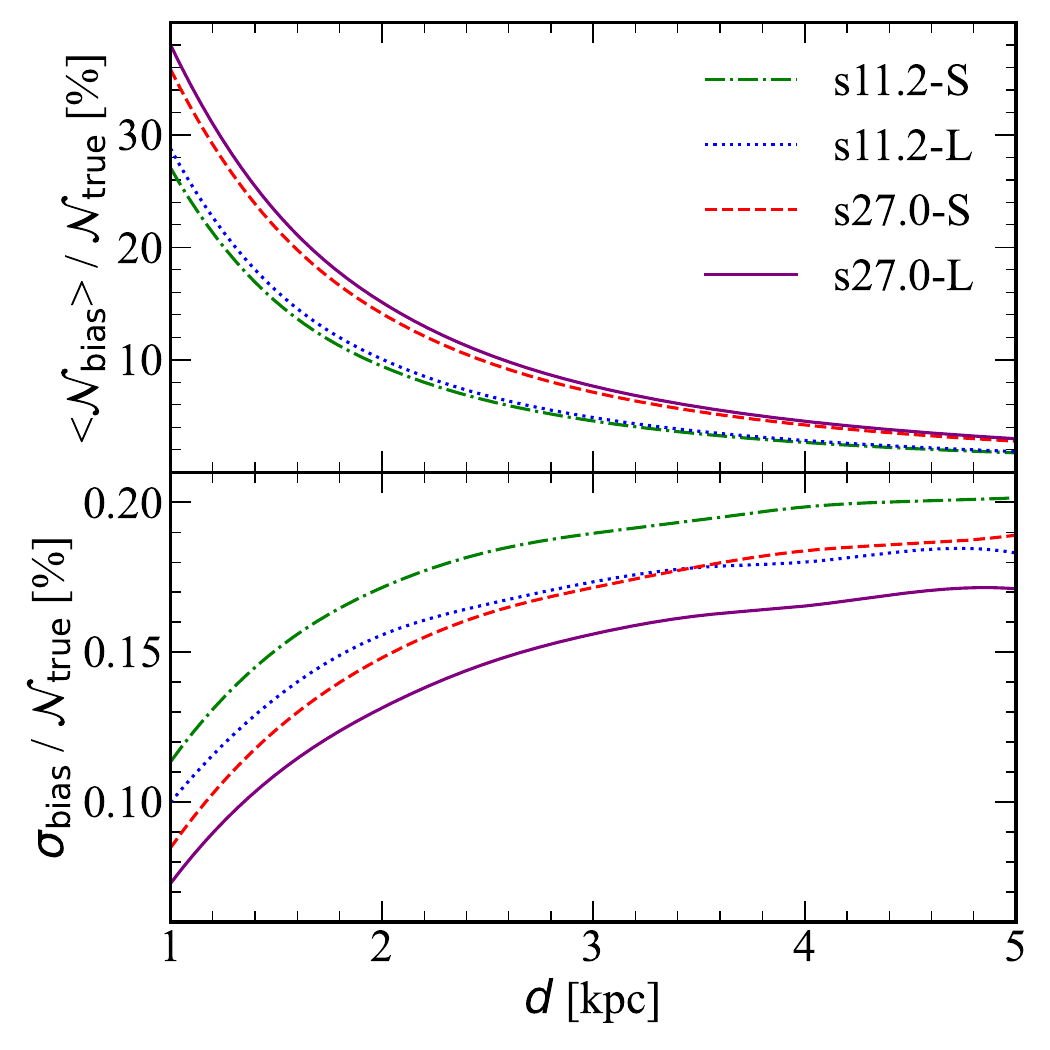}
    \caption{
    Simulated event pile-up in {Argo} with $\sin^2{\theta_W}=0.23863$.
    For each set of Monte Carlo simulations at a certain value of $d$, $3\times10^4$ individual mock triggers are generated.
    Upper panel: mean portion of pile-up events $\langle \mathcal{N}_\text{bias} \rangle/ \mathcal{N}_\text{true}$ (in percentage).
    Lower panel: $1\sigma$ variation of pile-up events $\sigma_{\text{bias}}/ \mathcal{N}_\text{true}$
    }
    \label{figure3}
\end{figure}

According to the study based on DarkSide-50~\cite{DarkSide:2018stg},
the observed S2 signal in a dual-phase LAr TPC is a pulse with a width of $\sim\!10\,\text{$\mu$s}$ (or a FWHM of $\sim\!5\,\text{$\mu$s}$).
When an intense neutrino flux from a nearby source comes into the LAr TPC, a large number of interactions occur within a short period of time, and some S2 pulses may overlap with each other.
Such a phenomenon will reduce the total number of observed events by a bias number $\mathcal{N}_\text{bias}$.
Meanwhile, $\mathcal{N}_\text{bias}$ will get a value randomly in the detection according to a probability distribution.
To evaluate $\mathcal{N}_\text{bias}$ and its uncertainty, we use the Monte Carlo method to study the pile-up issue, which was also used in dual-phase xenon TPCs in Ref.~\cite{Lang:2016zhv}.
We adopt the assumption that two S2 signals are distinguishable only if the spacing from the start of one pulse to the start of the next is larger than $10\,\text{$\mu$s}$.
Otherwise, these two close S2 signals will be combined as one.
The expected events are distributed randomly in time according to the scattering rate $\mathrm{d}\mathcal{N}/\mathrm{d}t$ and uniformly throughout the TPC.
Argo is assumed to have a height of 5\,m and a drift velocity of $0.93 \,\text{mm}/\text{$\mu$s}$~\cite{DarkSide20k:2020ymr}.
We also take into account the time delay due to the drifting of ionization electrons.
With the above setup, a set of Monte Carlo simulations is performed, which consists of $3\times10^4$ individual mock triggers for each $d$. After that, we count the number of unresolved events in each trigger and obtain a histogram of $\mathcal{N}_\text{bias}$ for this dataset.
Finally, the mean value $\langle \mathcal{N}_\text{bias} \rangle$ and uncertainty $\sigma_\text{bias}$ can be extracted from the histogram.
The Monte Carlo simulation results with $\sin^2{\theta_W}=0.23863$ are displayed in Fig.~\ref{figure3}.
As $d$ decreases, the mean portion of pile-up events $\langle \mathcal{N}_\text{bias} \rangle/ \mathcal{N}_\text{true}$ increases, but the relative $1\sigma$ variation $\sigma_{\text{bias}}/\mathcal{N}_\text{true}$ decreases.
This is generic for all 4 CCSN models. 
The magnitude of $\sigma_{\text{bias}}$ grows to $\sim\!100$ when $d$ approaches $1{\kpc}$, which is $\sim\!0.1\%$ of $\mathcal{N}_\text{true}$ in all models. 

The prediction of $\mathcal{N}_\text{bias}$ depends on the CCSN model (e.g., the progenitor mass) and $d$, while $\sigma_{\text{bias}}$ is less affected by these factors according to Fig.~\ref{figure3}. 
In a real measurement, the multi-messenger observation of a nearby CCSN will be used to calibrate the CCSN model, and also to measure $d$ (further discussions can be found in Sect.~\ref{Section 4.2}).
Hence, we only consider the distance dependence now.
We remark that, technically speaking, two S2 signals with a spacing smaller than $10 \,\text{$\mu$s}$ can still be distinguished if the S2 pulse shapes get read out in a real measurement~\cite{DarkSide:2018stg}. 
We also test the case of shrinking the criterion of resolving two successive events from $10 \,\text{$\mu$s}$ to $5 \,\text{$\mu$s}$ in the Monte Carlo simulations.
Using the new criterion, $\langle \mathcal{N}_\text{bias} \rangle/ \mathcal{N}_\text{true}$ is universally shifted down by a factor of $\sim\!1/2$ at $d\gtrsim3{\kpc}$ in all models, and by a factor of $\sim\!1/3$ at $d\sim1{\kpc}$. 
Another way to avoid a large portion of pile-up events is to only consider the cooling phase where the neutrino luminosity is less intense (e.g., $t\gtrsim1\,\text{s}$ for the models in Fig.~\ref{figure1}).
Our simulations find that the mean portions of pile-up events are reduced by $\sim\!50\%$ when only the cooling phases are considered.

Another issue is the uncertainty in the number of expected events in {Argo}. 
Although the time-dependent supernova neutrino fluxes are used to calculate the total number of expected counts in Eq.~\eqref{eq: chi-2 function} in this work, the time-integrated neutrino fluxes should be used in a real measurement. 
They are expected to be measured by large-scale neutrino observatories in parallel to {Argo}. 
That is, by analyzing the data from one detector or a combination of multiple detectors, parameters of the time-integrated neutrino spectra can be reconstructed via various methods (see, e.g., Refs.~\cite{Nikrant:2017nya,GalloRosso:2017mdz,Li:2017dbg,Li:2019qxi,Nagakura:2020bbw,Huang:2023aob}). 
We firstly estimate the uncertainty of the extracted supernova neutrino flux $\sigma_{\nu}$ (in percentage) in other detectors according to the parameter extraction precision of all flavors in a Bayesian approach~\cite{Huang:2023aob}.
The $2\sigma$ extraction precision for the observation of a $10{\kpc}$ target is presented in Table 3 of Ref.~\cite{Huang:2023aob}, for both normal mass ordering (NMO) and inverted mass ordering (IMO).
We first calculate the total neutrino spectra according to the $2\sigma$ parameter space and find the upper and lower boundaries. 
Then, we integrate the neutrino spectra to get the total numbers of neutrinos associated with the most probable spectrum as well as the two boundaries.
The $2\sigma$ flux uncertainty can thus be estimated as the difference in total numbers of neutrinos (in percentage) between the most probable spectrum and the two boundaries. 
We find that $\sigma_{\nu} \simeq 14.1\,(13.1)\%$ at $d=10{\kpc}$ assuming NMO (IMO). 
Furthermore, the authors of Ref.~\cite{Huang:2023aob} also tested the distance effect by assuming a closer source, e.g., $d=5{\kpc}$, and they find that the precisions are improved by about $40-50\%$ among almost all parameters.
Their results indicate that the reconstruction uncertainty is currently dominated by the statistics.
Consequently, we make such an assumption: $\sigma_\nu \propto 1/\sqrt{\mathcal{N}_\nu} \propto d$ since the total number of supernova neutrinos on the Earth is $\mathcal{N}_\nu \propto 1/d^2$. 
While $\mathcal{N}_{\text{true}}^{\text{th}} \propto \mathcal{N}_\nu$ and $\mathcal{N}_{\text{bias}}^{\text{th}}$ is also related to $\mathcal{N}_\nu$, we can estimate the flux uncertainty of expected events $\sigma_{\text{flux}}$ in {Argo} according to the uncertainty of the extracted supernova neutrino flux $\sigma_{\nu}$ in other detectors evaluated above. 
Thus, we adopt $\sigma_\text{flux} \simeq \mathcal{N}_{\text{obs}}^{\text{th}}\times14.1 \%\times(d/10{\kpc})$ assuming NMO.
To test such an estimate, we perform an additional set of Monte Carlo simulations for the model s27.0-L, following the aforementioned setting, and vary $\mathcal{N}_\text{true}$ by $\pm 10\%$. 
The resulting variation in $\mathcal{N}_{\text{obs}}$ is about $\pm 10\,(7)\%$ at $d=5\,(1){\kpc}$, smaller than the variation in $\mathcal{N}_\text{true}$ itself at small $d$. 
Therefore, we remark that the above expression of $\sigma_{\text{flux}}$ may slightly overestimate the flux extraction uncertainty for a very close source. 
The caveat is that the assumption of $\sigma_\nu \propto d$ may not work for a very close source, since other systematic errors may overwhelm the statistical error.

\begin{figure}[!ht]
    \includegraphics[width=\columnwidth]{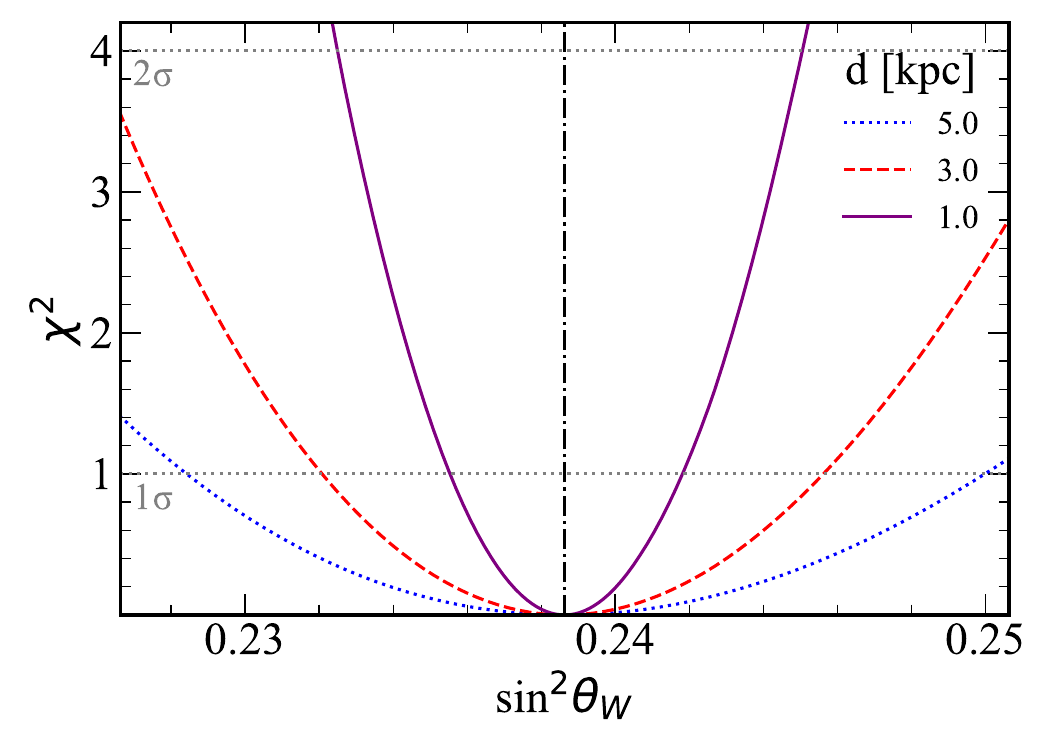}
    \caption{
    Expected sensitivity of {Argo} to the variation in low-energy $\sin{^2\theta_W}$ from the SM value of $0.23863$ (indicated by the vertical black dashed-dotted line), at a few benchmark values of $d$. 
    The $1\sigma$ and $2\sigma$ confidence levels are indicated by the horizontal grey dotted lines.
    The model s27.0-L and NMO are assumed
    }
    \label{figure4}
\end{figure}

Finally, we obtain the total systematic error of the low-energy $\sin{^2\theta_W}$ measurements by $\sigma^2_{\text{syst}} = \sigma^2_{\text{bias}} + \sigma^2_{\text{flux}}$. 
We evaluate the expected sensitivity of {Argo} to the variation in $\sin{^2\theta_W}$ using Eq.~\eqref{eq: chi-2 function}. 
The sensitivity curves at a few benchmark values of $d$ are shown in Fig.~\ref{figure4} for the model s27.0-L as an example.
NMO is assumed when evaluating $\sigma_{\text{flux}}$, and a similar result is obtained when IMO is assumed. 
Currently, $\sigma_{\text{flux}}$ is the most significant source of uncertainty for these benchmark values of $d$.
The expected precision of the low-energy $\sin{^2\theta_W}$ measurement is higher for a smaller $d$ due to the proportionality of $\sigma_{\text{flux}}$ to $d$.
In particular, a theoretical precision of $2.8\,(1.3)\%$ at the $1\sigma$ confidence level can be achieved at $d=3\,(1){\kpc}$, and it is $4.5\%$ at $d=5{\kpc}$.
Note that the corresponding curves for different CCSN models are nearly identical. 
We further remark that these results should generally be considered as upper limits of the precision due to incomplete uncertainty evaluation.

\subsection{Discussions} \label{Section 4.2}

\begin{figure}[!ht]
    \includegraphics[width=\columnwidth]{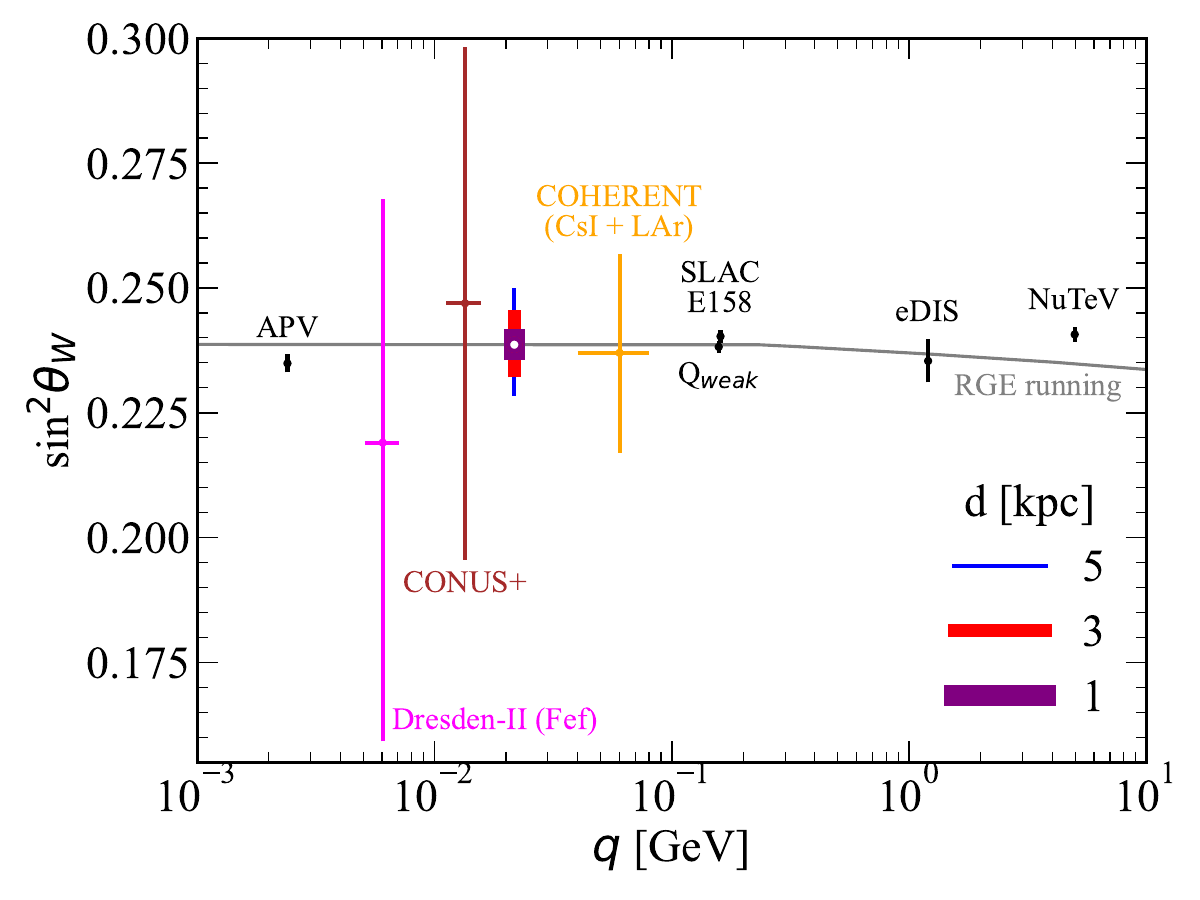}
    \caption{
    Comparison between the expected precision of low-energy $\sin{^2\theta_W}$ measurement by {Argo} 
    and other measurements~\cite{ParticleDataGroup:2024cfk}, at the $1\sigma$ confidence level, at different momentum transfers $q$. 
    Measurements from recent CE$\nu$NS experiments~\cite{AtzoriCorona:2022qrf,DeRomeri:2022twg,DeRomeri:2025csu} are also shown in colors
    }
    \label{figure5}
\end{figure}

Figure~\ref{figure5} shows the comparison between the expected $1\sigma$ precision of low-energy $\sin{^2\theta_W}$ measurement by {Argo}, as indicated in Fig.~\ref{figure4}, and the results of other experiments. 
It can be seen that the expected error bar is always smaller than those derived from recent CE$\nu$NS experiments, e.g., Dresden-II~\cite{AtzoriCorona:2022qrf}, COHERENT~\cite{DeRomeri:2022twg}, and CONUS+~\cite{DeRomeri:2025csu}.
Especially, we would like to compare our results with the expected sensitivities of the {CONNIE}, {MINER}, and {RED-100} reactor neutrino experiments~\cite{Canas:2018rng}, which also rely on the detection of CE$\nu$NS events. 
While the optimistic precision of these experiments is expected to be $\sim\!1\%$, a comparable precision ($\lesssim\!3\%$) may also be achievable for {Argo} if the supernova neutrino flux from a CCSN at $d\lesssim\!3{\kpc}$ is available.

The measurement of low-energy $\sin{^2\theta_W}$ also sheds light on searching for new physics.
For example, the interactions between low-energy neutrinos and nucleons could potentially involve new physics beyond the SM, which can be described by the neutrino NSIs~\cite{Breso-Pla:2023tnz,Cadeddu:2023tkp}. 
As a result, these NSIs could introduce an effective shift in the SM value of $\sin{^2\theta_W}$, which can be identified by {Argo} due to the modification in the differential cross section of CE$\nu$NS. 
By considering the neutrino NSIs with $u$ and $d$ quarks described by the effective Lagrangian~\cite{Davidson:2003ha,Barranco:2005yy,Scholberg:2005qs}, the weak charge $Q_W$ and the weak form factor of $^{40}\text{Ar}$ are modified~\cite{Coloma:2017egw}.
Based on the anticipated high-precision measurements of $\sin{^2\theta_W}$ at low energies, as illustrated in Fig.~\ref{figure4}, the sensitivity to the NSI parameters, such as $\varepsilon_{\alpha \alpha}^{u V}$ and $\varepsilon_{\alpha \alpha}^{d V}$, could reach the range of (-0.006, 0.006) at the $90\%$ confidence level ($\chi^2 = 2.71$) for a CCSN event located at $d\!\sim\!3{\kpc}$. 
Such an improvement in sensitivity enhances the precision of constraining the neutrino NSI parameters by approximately one order of magnitude compared to the first COHERENT data~\cite{Papoulias:2017qdn, Coloma:2017ncl, Khan:2019cvi, Papoulias:2019txv, Giunti:2019xpr}.

Note that the uncertainty of distance is not included yet. 
Currently, the distances of some candidate pre-supernova stars can be measured with a precision of a few percent ($<5\%$) in Gaia missions~\cite{2023A&A...674A...1G,Mukhopadhyay:2020ubs,2018A&A...616A...1G}. 
Recent research finds that distance estimate with neutrinos can also achieve similar precision~\cite{Healy:2023ovi}. 
It is reasonable to expect that a dedicated multi-messenger observation in the future can achieve higher precision.
Here we make a simple estimate of the uncertainty originating from the uncertainty of distance. 
We take $\mathcal{N}_{\text{obs}}$ at $d=3{\kpc}$ as the observed data, then we vary $d$ by $\pm1\%$ when calculating the theoretical results and uncertainties. 
The combined $1\sigma$ precision of $\sin{^2\theta_W}$ changes to $4.1\%$ from $2.8\%$ without considering the distance uncertainty.

We remark that the supernova neutrino flux is important in our evaluation.
When we evaluate the total number of CE$\nu$NS events, we can simply use the time-integrated neutrino flux reconstructed from other observations, and its uncertainty is considered in $\sigma_{\text{flux}}$.
The time-dependent neutrino flux from a well-calibrated supernova simulation would be necessary for the estimation of pile-up events.
So, the uncertainty in the CCSN model has less influence on the final results since it only affects the estimate of pile-up events in a real measurement.
In this work, nevertheless, we adopt 4 CCSN models to estimate the event pile-up in {Argo} detector.
The results (see Fig.~\ref{figure3}) indicate that the number of unresolved events depends on the progenitor mass, nuclear EoS, and source distance.
Unlike the progenitor mass and source distance that will be measured via multi-messenger observations for a nearby target~\cite{Mukhopadhyay:2020ubs,Pledger:2023ick,2023A&A...674A...1G,Healy:2023ovi,2018A&A...616A...1G}, the nuclear EoS properties remain uncertain and require further determinations~\cite{FiorellaBurgio:2018dga,Burgio:2021vgk,Lattimer:2021emm}.
For example, the value of nuclear matter incompressibility is reported to be $240\pm20{\MeV}$ from isoscalar giant monopole resonance experiment, and is adopted in various relevant studies~\cite{Youngblood:1999zza,Shlomo:2006ole,Chen:2011ps,Colo:2013yta,Garg:2018uam,Cai:2021ucx,Li:2021thg}. 
The incompressibility of the LS220 and Shen EoS, employed by these CCSN models, is $220$ and $281{\MeV}$, respectively~\cite{Lattimer:1991nc,Shen:1998gq}.
Such choices represent two extremes with respective to the aforementioned experimental constraint.
The difference in pile-up portion $\langle \mathcal{N}_\text{bias} \rangle/ \mathcal{N}_\text{true}$ between the CCSN models using the LS220 and Shen EoSs increases from $0.11\,(0.22)\%$ at $d=5{\kpc}$ to $1.73 (2.19)\%$ at $d=1{\kpc}$ for the models with progenitor mass being $11.2\, (27.0)$ M$_\odot$ (see Fig.~\ref{figure3}). 
Hence, we believe that the influence of the nuclear EoS uncertainties on our results may be noticeable at a very small $d$, and further analysis is required to study its impacts quantitatively.
Note that, however, different values and confidence ranges of incompressibility are obtained using other approaches (see, e.g., Refs.~\cite{Leifels:2015iei,Huth:2021bsp,Sorensen:2023zkk}). 
It is worth paying attention to the latest progress on constraining nuclear EoS through astrophysical observations and nuclear experiments~\cite{Raaijmakers:2021uju,Zhang:2021xdt,Pang:2021jta,Radice:2017lry,LIGOScientific:2018hze}.
We expect an even smaller uncertainty from the nuclear EoS once the supernova simulation is further calibrated by future observations and experiments.

Another issue is the occurrence of the next galactic CCSN.
A list of 31 CCSN candidates within $1{\kpc}$ from the Earth is suggested in Ref.~\cite{Mukhopadhyay:2020ubs}.
More candidates can be expected within a range of $5{\kpc}$.
Meanwhile, the supernova rate is estimated to be $1.63\pm0.46$ per century for the Milky Way Galaxy~\cite{Rozwadowska:2020nab}. 
Nonetheless, it has been 420 years since the last observed galactic supernova, i.e., the Kepler supernova in 1604~\cite{Bethe:1990mw}, and we can tell that no galactic supernova has occurred for at least $38$ years, according to the neutrino detection, since the extra-galactic SN 1987A. 
Among the last 1,000 years, only 6 galactic supernovae were recorded: Lupus at $2.2{\kpc}$ in 1006, Crab at $2.0{\kpc}$ in 1054, 3C 58 at $2.6{\kpc}$ in 1181, Tycho at $2.4{\kpc}$ in 1572, Kepler at $4.2{\kpc}$ in 1604, and Cas A at $2.92{\kpc}$ in 1680, all happened within $5{\kpc}$ from the Earth. 
$4$ of these events out of $6$, namely Crab, 3C 58, Kepler, and Cas A, were considered to be CCSNe~\cite{The:2006iu}. 
The next galactic supernova seems to be imminent and possible to be a CCSN.  
It is meaningful to get prepared to catch such an once-in-a-life-time chance. 

\section{Conclusions} \label{Section 5}

We study the low-energy $\sin{^2\theta_W}$ measurement by {Argo} in the observation of a nearby CCSN explosion, through the CE$\nu$NS interaction.
A galactic CCSN event provides an intense neutrino flux with a mean energy of $\sim10-20{\MeV}$, and will be observed by both {Argo} and other dedicated neutrino observatories.
Given the great progress in neutrino detection in the last decades, the next nearby CCSN will be well observed and also be used to calibrate the CCSN simulations. 
The {Argo} detector can collect the CE$\nu$NS signals when those MeV neutrinos scatter off the argon nuclei, providing a way to extract the value of low-energy $\sin^2{\theta_W}$. 
We further consider the systematic uncertainties due to the flux extraction of supernova neutrinos and event pile-up, and find that a precision of $\lesssim\!3\%$ is potentially possible at a supernova distance $d\!\lesssim\!3{\kpc}$.
Such a precision is better than the results from recent CE$\nu$NS experiments, and comparable to the optimistic expected sensitivity of ongoing experiments (e.g., the {CONNIE}, {MINER}, and {RED-100} reactor neutrino experiments).
Our analysis of the systematic uncertainties can provide valuable information for the experimentalists to improve their strategy in the observation of the next nearby CCSN.
However, the caveat is that this result should be taken as an upper limit of the precision due to the currently incomplete uncertainty estimation. 
We leave the detailed analysis of the systematic uncertainties in a real measurement to future work.

The precise measurement of the low-energy $\sin{^2\theta_W}$ is important for the precision test of the SM.
Any deviation may provide a hint to new physics, especially in the neutrino sector where the existence of new physics has already been proven by neutrino oscillation experiments.
Specifically, such a measurement can be used to constrain the neutrino NSI parameters.
Recently, the PandaX~\cite{PandaX:2024muv} and XENON~\cite{XENON:2024ijk} collaborations reported the first detection of solar $^{8}\text{B}$ neutrinos by direct DM detectors, through the CE$\nu$NS channel.
Since the solar $^{8}\text{B}$ neutrinos have been well-studied, future multi-ton scale DM detectors may also provide excellent opportunities to determine the low-energy $\sin{^2\theta_W}$ via such measurements.

\bmhead{Acknowledgements}

We acknowledge Hans-Thomas Janka, Daniel Kresse, Lorenz Hüdepohl, and Alessandro Mirizzi for granting access to the Garching Core-Collapse Supernova Archive (\url{https://wwwmpa.mpa-garching.mpg.de/ccsnarchive}). 
We thank Tianyu Zhu and Kaixuan Ni for the discussions on DM detectors.
This work is partially supported by grants from the Research Grant Council of the Hong Kong Special Administrative Region, China (Project Nos. 14300320 and 14304322). 
This material is based upon work supported by the National Science Foundation under Grant AST-2316807.

\bibliography{sn-bibliography}

\end{document}